\documentclass[conference]{IEEEtran}
\IEEEoverridecommandlockouts
\usepackage{float}
\usepackage{cite}
\usepackage{amsmath,amssymb,amsfonts}
\usepackage{algorithmic}
\usepackage{graphicx}
\usepackage{textcomp}
\usepackage{placeins}
\usepackage[dvipsnames]{xcolor}
\usepackage[a4paper, total={184mm,239mm}]{geometry}
\usepackage{svg}
\usepackage[final, activate={true, nocompatibility}, kerning=true, babel=true]{microtype}

\usepackage[pscoord]{eso-pic}

\def\BibTeX{{\rm B\kern-.05em{\sc i\kern-.025em b}\kern-.08em
\kern-.1667em\lower.7ex\hbox{E}\kern-.125emX}}

\begin{document}
% \vspace{-1em}
\title{SALSA: \underline{S}imulated \underline{A}nnealing based \underline{L}oop-Ordering \underline{S}cheduler for DNN \underline{A}ccelerators}

% \placetextbox{0.5}{1}{\footnotesize This work has been submitted to the IEEE for possible publication. Copyright may be transferred without notice, after which this version may no longer be accessible.}

\author{
\thanks{This work is funded in part by the Convolve project evaluated by the EU Horizon Europe research and innovation program under grant agreement No. 101070374 and has been supported by the Swiss State Secretariat for Education Research and Innovation under contract number 22.00150. E-mail: \{jungvi / lbenini\}@iis.ee.ethz.ch \{arne.symons / linyan.mei / marian.verhelst\}@esat.kuleuven.be}
\IEEEauthorblockN{Victor J.B. Jung$^\dagger$, Arne Symons$^*$, Linyan Mei$^*$, Marian Verhelst$^*$, Luca Benini$^\dagger$}
\IEEEauthorblockA{$^\dagger$Integrated Systems Laboratory, ETH Zürich, Switzerland. $^*$Department of Electrical Engineering, KU Leuven, Belgium.}}

\maketitle
\vspace{-1em}

\begin{abstract}

To meet the growing need for computational power for DNNs, multiple specialized hardware architectures have been proposed. Each DNN layer should be mapped onto the hardware with the most efficient schedule, however, SotA schedulers struggle to consistently provide optimum schedules in a reasonable time across all DNN-HW combinations. 

This paper proposes SALSA, a fast dual-engine scheduler to generate optimal execution schedules for both even and uneven mapping. We introduce a new strategy, combining exhaustive search with simulated annealing to address the dynamic nature of the loop ordering design space size across layers. SALSA is extensively benchmarked against two SotA schedulers, LOMA \cite{symons2021loma} and Timeloop \cite{parashar2019timeloop} on 5 different DNNs, on average SALSA finds schedules with 11.9\% and 7.6\% lower energy while speeding-up the search by 1.7$\times$ and 24$\times$ compared to LOMA and Timeloop, respectively.

\end{abstract}


\vspace{-0.5em}
\section{Introduction}

Convolutional Neural Networks (CNNs) \cite{lecun1995convolutional} are a very successful class of machine learning (ML) models. This type of Deep Neural Network (DNN) consists of a stack of convolutional layers and reaches state-of-the-art (SotA) performance in the fields of image recognition, classification, segmentation, etc. A wide range of specialized hardware (HW) emerged to accelerate DNN execution \cite{chen2019eyeriss}. These DNN accelerators vary from datacenter-class \cite{joupi2017tpu} to embedded systems. The efficiency of a DNN Accelerator is mainly based on the memory hierarchy, the spatial unrolling, and it heavily relies on efficient schedulers to find optimal temporal mappings \cite{sze2017efficient} of DNN layers onto hardware resources.

As previous work has demonstrated, the scheduling of a NN onto such HW accelerators impacts energy and latency up to orders of magnitude \cite{kwon2019understanding}.
A subtle change in the characteristics of the NN-HW combination can completely modify the optimal schedule. For example, a change in on-chip memory resources can alter the optimal data allocation scheme and even the most efficient workload execution order to minimize energy or latency. 

As a result, many design space exploration (DSE) schedulers \cite{parashar2019timeloop}, \cite{yang2020interstellar}, \cite{huang2021cosa}, \cite{symons2021loma}, have been proposed to automatically find the optimal schedule given a DNN workload and an accelerator HW architecture. 
However, the above-mentioned schedulers fail to reach near-optimal mappings in a reasonable time. 
The contributions of this paper are the following:

\begin{enumerate}
    \item{We introduce SALSA, a novel scheduler that never shrinks or prunes the schedule search space while having an execution time of a few seconds. Using a dual-engine strategy, SALSA consistently reaches near-optimal schedules with an average error margin of $0.007$\%.}
    \item{To prove its superiority, we extensively compare SALSA with 2 SotA schedulers, LOMA \cite{symons2021loma} and Timeloop \cite{parashar2019timeloop}. SALSA always finds schedules with higher or equal quality than Timeloop and LOMA while consequently reducing the search time.}
\end{enumerate}

We tested SALSA on 5 commonly used DNNs, benchmarked against Timeloop and LOMA, and evaluated using the SotA cost model ZigZag \cite{mei2020zigzag}. 
\textbf{
In both benchmarks, SALSA achieves a consequent reduction of the search time, we report 1.7$\times$ and 24$\times$ faster search than LOMA and Timeloop.
Most importantly, SALSA reaches superior schedules leading to a reduction of the energy needed to execute the model by 7.6\% and 11.9\% compared to LOMA and Timeloop, respectively.}

\begin{figure}[t]
\centering
\includegraphics[width=\columnwidth]{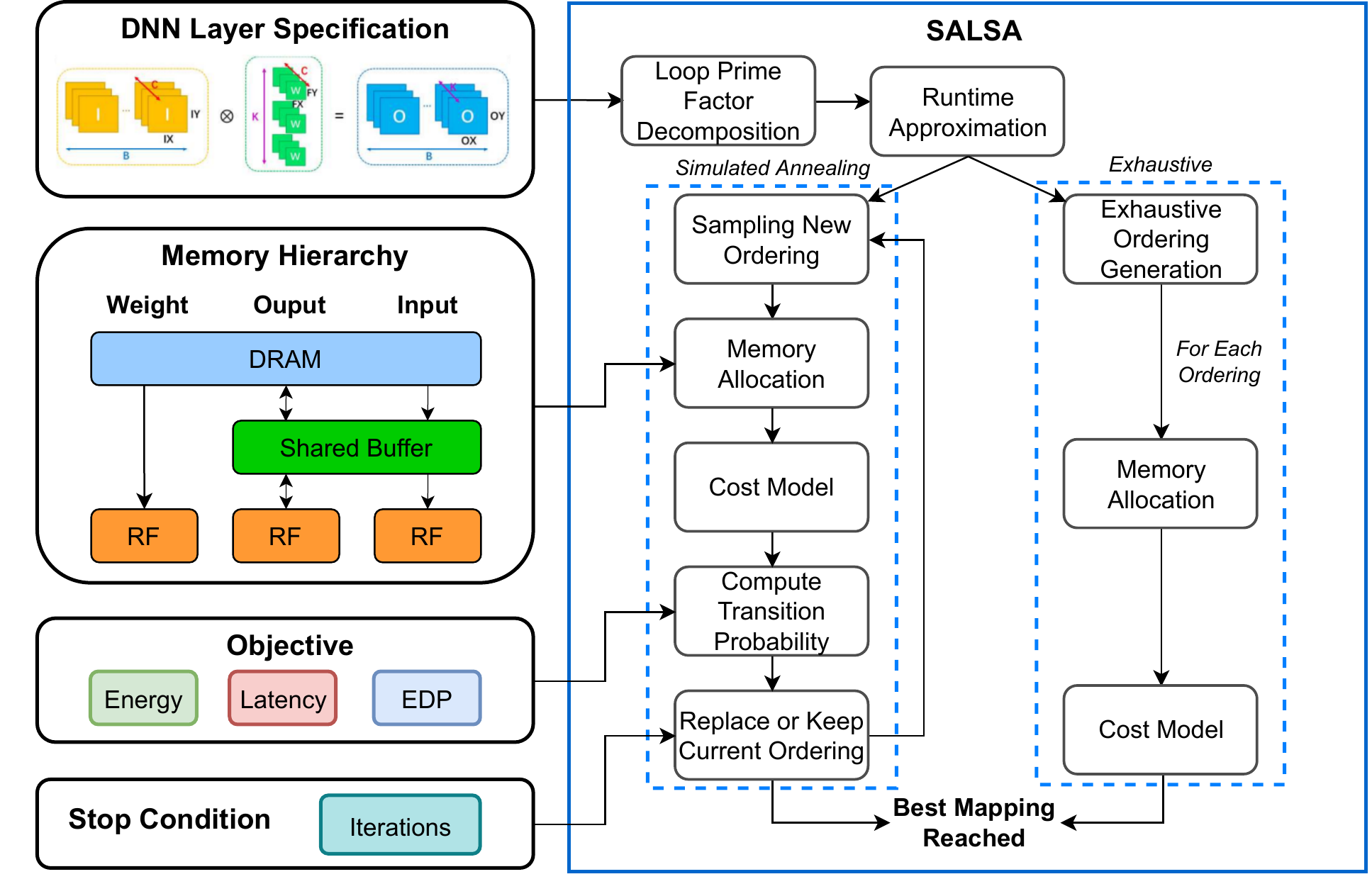}
\caption{Overview of the SALSA implementation.}
\label{fig:salsaoverview}
\vspace{-1.25em}
\end{figure}

\vspace{-1em}
\section{Background}
\label{sec:background}

    \subsection{DNNs, Accelerators \& Schedules}
    \label{subsec:wasm}
    
    A single convolutional layer consists of 7 nested for-loops, as can be seen in the top-left of Figure \ref{fig:salsaoverview}. The loop dimension sizes determine the tensor size of the three operands; Input~(I), Weight~(W), and Output~(O). Other NN layer topologies (fully connected, pointwise convolutional, matrix-matrix multiplication, etc.) can use the same representation by fixing specific loop dimension sizes to 1. In order to speed up the DNN inference or increase its energy efficiency, various Application-Specific Integrated Circuit (ASIC) DNN accelerators have been proposed both in academia and by the industry.
    Such accelerators typically include a spatially unrolled array of Processing Elements (PE) that consist of a Multiply-Accumulate (MAC) unit and local memories to store the operand data. The PEs are connected to larger memories higher up in the memory hierarchy stack through fixed interconnections or a flexible Network-on-Chip (NoC) \cite{chen2019eyeriss}. These connections allow the multicasting of operand data to multiple PEs, consequently parallelizing the computation.
    Unrolling a for-loop onto multiple PEs will turn it into a parallel for-loop (parfor-loop). 
    When executing a DNN onto an Accelerator, the set of parfor-loops is named spatial unrolling and indicates the parallelization pattern.
    Usually, the number of PEs is lower than the dimension of the original for-loops; thus, it is common to split them in order to turn a part of the original for-loop into parfor-loops.
    
    On top of the spatial unrolling, an optimized temporal execution schedule is crucial to extract the full potential of DNN Accelerators. More specifically, a schedule can be decomposed into two elements: 1.) the \textit{loop ordering}, which describes the temporal processing order of the for-loops, and 2.) the \textit{memory allocation}, which assigns the operands of each loop to a specific memory resource. A detailed description of these elements follows later.
    
    \subsection{Loop Prime Factor Decomposition}
    \label{subsec:lpfdecomposition}
    
    The loop ordering of the original nested for-loop representation would not result in an optimal schedule. By decomposing the large loop dimensions into multiple smaller loops, and subsequently re-ordering those smaller loops, better schedules can be found. At the finest level of granularity, each loop is decomposed into the number of prime factors of its loop dimension. The resulting indivisible for-loops are referred to as Loop Prime Factors (LPF). An example of the decomposition of an originally nested for-loop to an LPF ordering is shown in Fig.\ref{fig:saoverview} steps A to B.
    
    \subsection{Loop Ordering Search Space}
    \label{subsec:losearchspace}
    
    A loop ordering $o$ can be seen as a permutation of a finite set of elements, where each element represents a for-loop (Fig.\ref{fig:saoverview} step B). 
    The loop ordering search space is thus represented by the Symmetric Group $S_n$ with $n$ the number of loops in $o$. The order (number of elements) of $S_n$ is $n!$ if every element is unique. Due to the LPF decomposition, $n=20$ is not uncommon for modern DNN layers. This would require the evaluation of $O(10^{18})$ orderings. Therefore, exhaustively going through all elements in $S_n$ is only tractable for small NN layers where $n < 11$.

    \vspace{-0.5em}
    \subsection{Memory Allocation}
    \label{subsec:memalloc}
    
\begin{figure}[htbp]
    \centering
    \vspace{-0.5em}
    \includegraphics[width=\columnwidth]{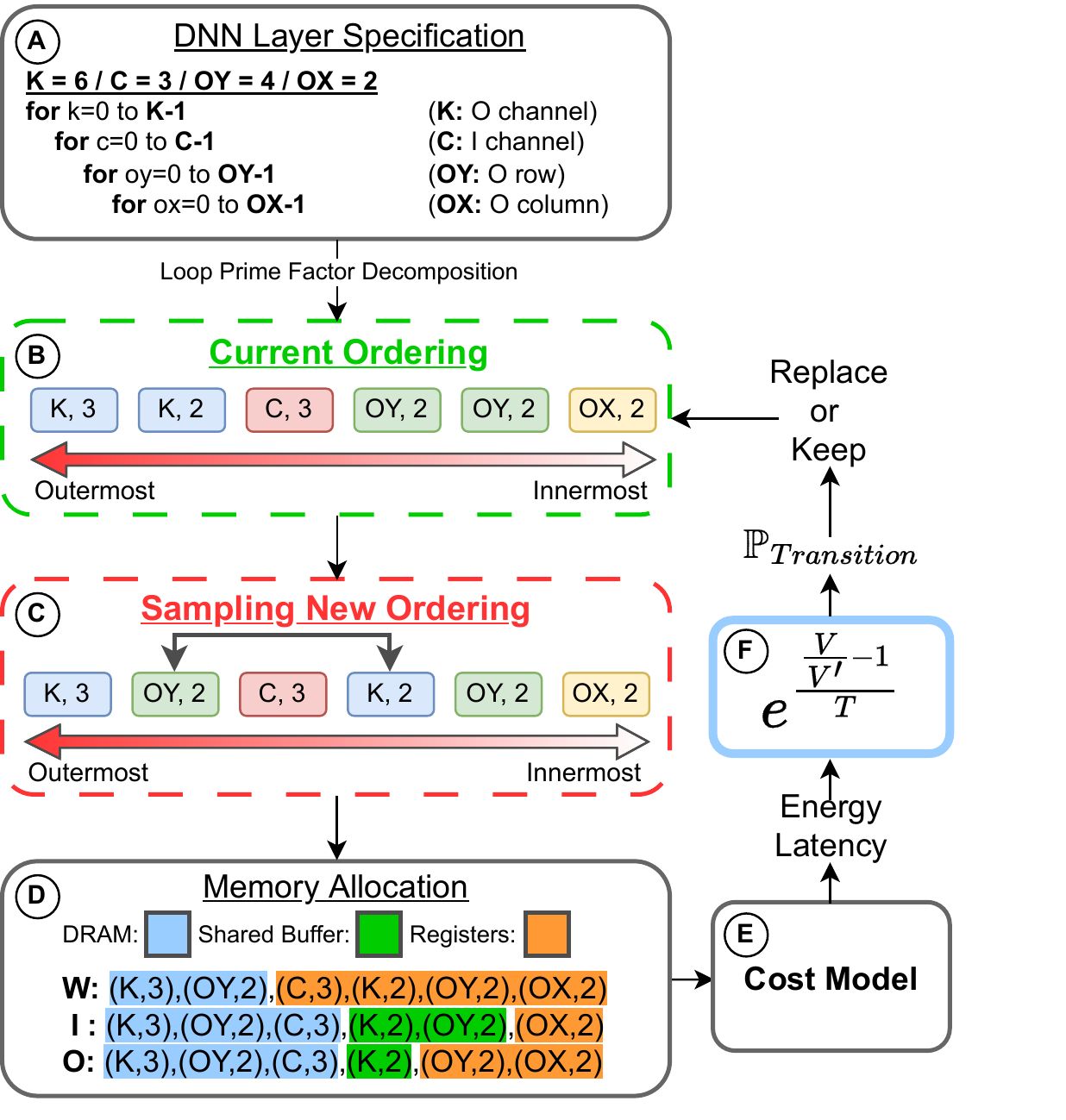}
    \vspace{-2em}
    \caption{Detailed example of SALSA's Simulated Annealing path. The workload used in this figure is fictional for the purpose of demonstration, and the Memory Hierarchy is composed of three levels: DRAM, Shared Buffer, and Registers.}
    \label{fig:saoverview}
    \vspace{-0em}
    \end{figure}

    Loop ordering has to be combined with the allocation of the data attributed to these loops to specific memory resources in the memory hierarchy (Fig.\ref{fig:saoverview} step D).
    Most mapping representations store the 3 operands (I/W/O) associated with a for-loop at the same memory level. Such mappings are referred to as `even memory mappings'. 
    A more complex mapping strategy has been proposed recently \cite{mei2020zigzag}, named 'uneven memory mapping'.
    This strategy allows to unevenly distribute of operand data of the nested for-loops within shared memories in the hierarchy in order to more efficiently exploit the data reuse at the cost of drastically enlarging the mapping search space. 
        
    To reduce the large mapping search space, LOMA \cite{symons2021loma} proposed a bottom-up memory allocation strategy independent of the loop ordering. This is possible due to the fact that for a single loop ordering $o$, the optimal memory allocation $m$ can be inferred with a one-to-one relationship in a bottom-up fashion.
    
    \subsection{Cost Model}
    
    The energy, latency, or any other performance metric of the inference of a CNN layer on an accelerator depends on four aspects: 1.) the DNN workload $w$ (size of the 7 loop dimensions); 2.) the accelerator characteristics $a$ (PE array size, memory organization, memory size, etc.); 3.) the spatial unrolling $s$ (parallelization strategy across PE array); 4.) the schedule or temporal mapping $m$.
    
    This work focuses on temporal DNN mapping optimization, where the inputs $w$, $a$, and $s$ are provided by the user or by an upper-level search engine. 

    The optimization objective, returned by the cost model, is noted $V$ and can represent the energy, latency, Energy-Delay Product (EDP), etc.
    
\vspace{-1.5em}
\section{Related Work}
\label{sec:relatedwork}
    
    In recent years, a plethora of tools has been proposed to generate high-quality schedules. Some constrain the search space like CoSA \cite{huang2021cosa}, and Pluto \cite{bondhugula2016pluto+} to speed up the search. Others, like Interstellar \cite{yang2020interstellar} and ZigZag \cite{mei2020zigzag} prune some part of the search space during the search through heuristics. LOMA \cite{symons2021loma} combines an exhaustive search with optional user-provided constraints, providing both unconstrained and constrained search. 
    Timeloop \cite{parashar2019timeloop} embeds a random search engine in an unconstrained space, failing to consistently provide near-optimum schedules in fast search time.
    Alternatively, Mind Mappings \cite{hegde2021mind} trains a DNN to substitute the cost model and make the search space smooth and differentiable in order to apply Stochastic Gradient Descent. 
    
    It is important to highlight that, besides the search strategy, the representation of a schedule varies between frameworks. This makes it hard to extensively compare results and performances. All the above-mentioned frameworks implement an even mapping representation (see Section~\ref{subsec:memalloc}). ZigZag's and LOMA's representation also allows uneven mappings. Consequently, its mapping search space becomes more complex.
    
    SALSA overcomes these bottlenecks by implementing a flexible and fast scheduler that allows for both even and uneven mappings generation by separating loop ordering and loop memory allocation in two independent processes. This also allows one to use SALSA with other scheduling representations, e.g., plug in another memory allocation strategy or cost model. As SALSA's loop ordering algorithm doesn't use expert knowledge of the cost model or memory allocation, it is robust to drastic changes in the search space.

\section{SALSA Scheduling Approach}
\label{sec:salsaschedulingapproach}
    
    To cope with the changing size of the search space from one layer to another, SALSA implements a dual search strategy, as shown in Figure~\ref{fig:salsaoverview}. The simulated annealing path is shown in detail in Figure~\ref{fig:saoverview}. 
    
    \subsection{Runtime Approximation and Search Method Selection}

    To decide which of the Exhaustive or Simulated Annealing paths is the fastest (Fig.\ref{fig:salsaoverview}), we evaluate and compare their runtime. The Simulated Annealing path's runtime is constant (depends on a fixed hyperparameter) while the Exhaustive path's execution time $T$ is evaluated as follows: 
    \begin{equation}
        T(n, k) = \tau \frac{n!}{\prod_{i=1}^{m} k_{i}!}
    \end{equation}
    Where $n$ is the number of elements in the loop ordering, $k_i$ is the multiplicity of the i-th element, $m$ is the number of unique elements in the loop ordering, and $\tau$ is an HW-dependent constant.
    
\begin{figure}[htbp]
    \centering
    \vspace{-0.5em}
    \includegraphics[width=\columnwidth]{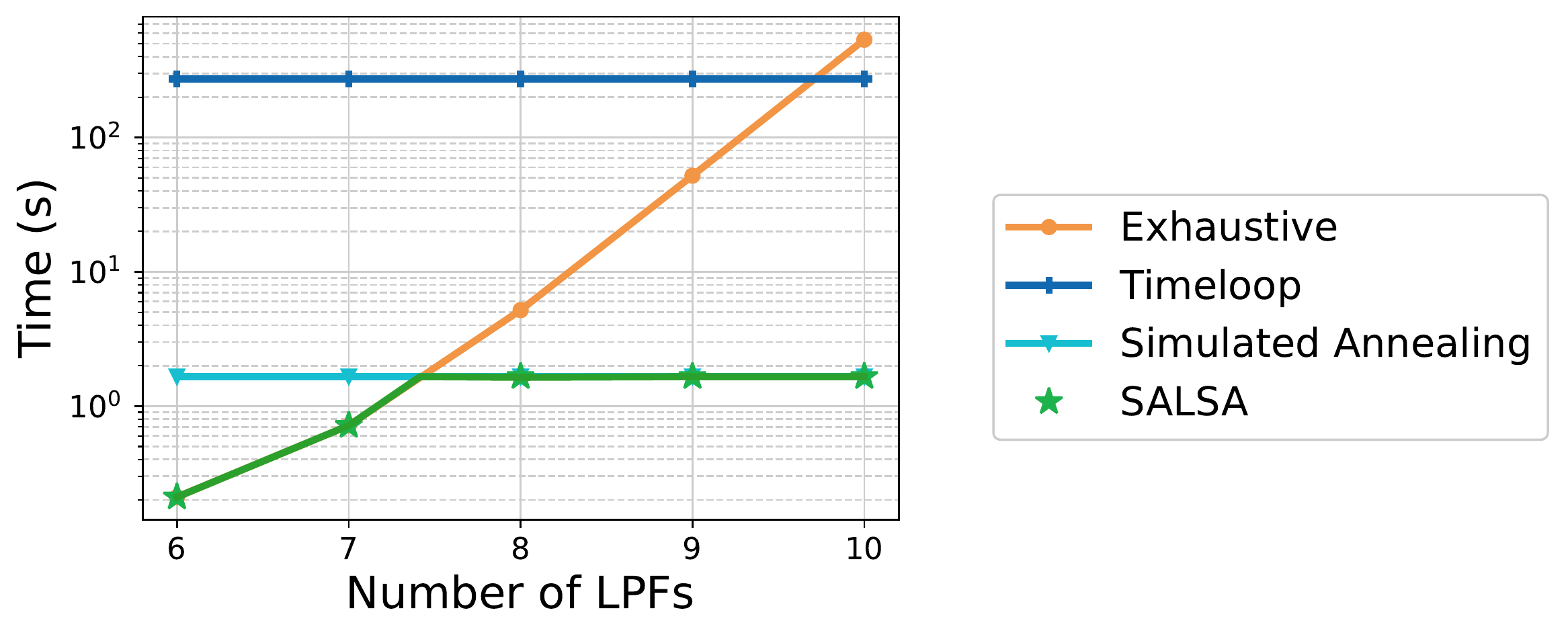}
    \vspace{-1.5em}
    \caption{Graph illustrating the required search time for different search strategies for varying numbers of LPFs for AlexNet Layer 2. Note the logarithmic y-axis.}
    \label{fig:exectime}
    \vspace{-1.5em}
    \end{figure}

    Figure~\ref{fig:exectime} shows how the exhaustive search time exponentially increases with the number of LPFs in a loop ordering while the simulated annealing search time remains constant.
    We will demonstrate that, even though more LPFs imply a larger permutation space, simulated annealing performs well across all DNN-HW combinations in a constant time.
    
    \subsection{Exhaustive Search}

    The exhaustive search branch is implemented using LOMA's scheduler \cite{symons2021loma}. After the exhaustive loop ordering generation, each unique ordering undergoes a bottom-up memory allocation and, finally, a cost model evaluation (both explained next). Most importantly, this exhaustive search engine guarantees to find the global optimum for any preferred optimization criterion at the cost of a potentially infeasible search time.

    \vspace{-1em}
    \subsection{Simulated Annealing Search}
    \label{subsec:simulatedannealing}
    
    In most cases, the exhaustive path would be too time-consuming, and thus the simulated annealing path is taken. Despite its simplicity, simulated annealing \cite{kirkpatrick1983optimization} and its different variants are widely used and prove to be efficient in combinatorial optimization. 
    Each iteration of the simulated annealing pass will go through the subsequent steps depicted in Figure~\ref{fig:saoverview}:
    
    \subsubsection{Sampling New Ordering}
    
    In order to sample new orderings (Fig.\ref{fig:saoverview} step C), we model a neighborhood of nearby states that can serve as the next candidate state \cite{kirkpatrick1983optimization}. SALSA defines the neighborhood of a loop ordering $o$ as follows:
    \begin{equation}
        N_{o} := \{ swap(o,i,j) \mid i \in [0,n),\ j \in [0,n),\ i \neq j \}
    \end{equation}
    with $swap(o,i,j)$ the action of swapping the LPFs at indices $i$ and $j$ of the ordering $o$ of size $n$. With this neighborhood, any point in the search space can be reached in $n-1$ steps. 
    
    \subsubsection{Memory Allocation \& Cost Model Evaluation}

    Firstly, we allocate the memory accordingly to the new loop ordering generated by the previous stage (Fig.\ref{fig:saoverview} step D). SALSA then uses a cost model to get the performances associated with the candidate state (Fig.\ref{fig:saoverview} step E). In this paper, results using the ZigZag as well as the Timeloop cost model will be shown.
    
    \subsubsection{Transition Probability Computation \& Next Node Selection}
    
    Once the cost $V'$ of the sampled state $m'$ is returned by the cost model, SALSA computes the probability of accepting the candidate state $m'$ using the following formula:

    \vspace{-0.5em}
    \begin{equation}
        \mathbb{P}(m, m') = exp(\frac{\frac{V}{V'}-1}{T})   
    \end{equation}
    
    where $V$ and $V'$ are respectively the optimization objective of the states $m$ and $m'$. The temperature $T$ is a hyperparameter handling the balance between \textit{intensification} and \textit{diversification} to avoid getting stuck in local optima while focusing the search on promising regions of the search space.
    
    The evolution of $T$ depends on the number of iterations $I$ and respect the following geometric progression: $T_{i+1}=\rho T_i$ where $\rho = 0.999$

    \begin{figure}[H]
    \vspace{-0.5em}
    \includegraphics[width=\columnwidth]{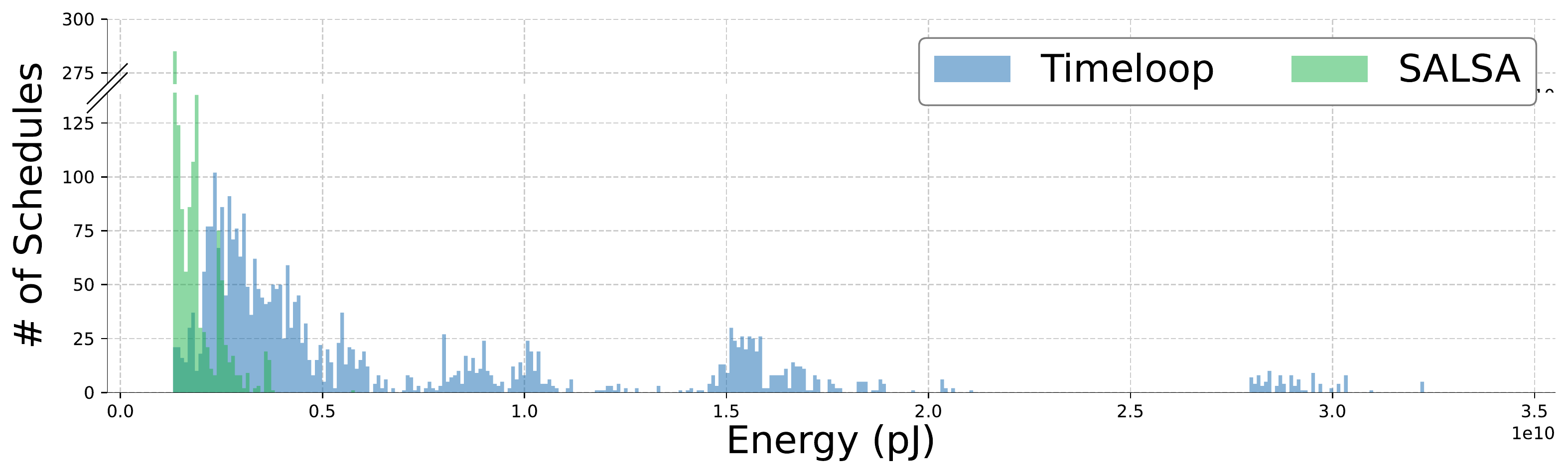}
    \vspace{-1.5em}
    \caption{Mapping energy distribution during a search for layer 2 of AlexNet. using Timeloop and SALSA. Best viewed in color.}
    \vspace{-1em}
    \label{fig:energydistrib}
    \end{figure}

    \begin{figure*}[ht]
    \centering
    \includegraphics[ width=\textwidth]{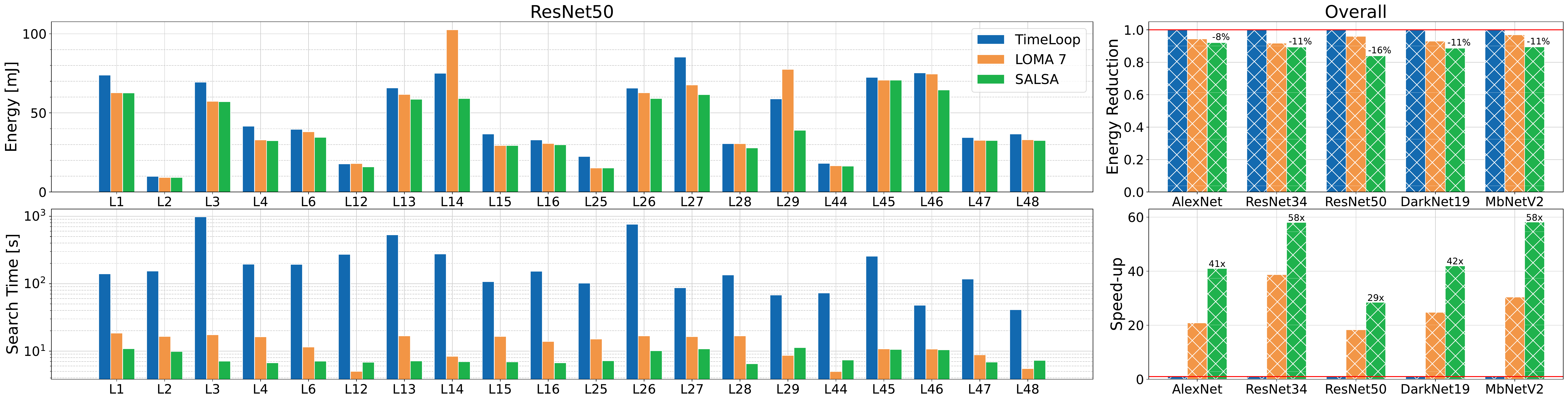}
    \vspace{-2em}
    \caption{Comparison of SALSA, LOMA 7, and Timeloop for 5 DNN. In this figure, LOMA is configured with an LFP limitation factor of 7. The left part displays the Energy and Search Time for every unique layer of ResNet50, while the right part shows the average Energy Reduction and Speed-up of each DNN. Energy Reduction and Speed-up in the right plots are normalized with Timeloop's Energy and Time, respectively.}
    \vspace{-1em}
    \label{fig:unevenmerged}
    \end{figure*}  
\vspace{-1em}
\section{Experimental Results and Benchmarking}
\label{sec:benchmark}

    \subsection{Experimental Setup}
    
    SALSA is implemented in Python and benchmarked across other schedulers available in the SotA. In our study, we use the 5 following NN: AlexNet, ResNet34 \cite{he2016deep}, ResNet50 \cite{he2016deep}, DarkNet19 \cite{redmon2013darknet}, and MobileNetV2 \cite{mbnetv2}. The accelerator $a$ is an Eyeriss-like architecture \cite{chen2019eyeriss}, consisting of a 14 by 12 PE array. Besides a MAC unit, each PE includes a scratchpad for weights, inputs, and outputs. Above the PE array resides a global buffer for storing inputs and outputs, followed by a DRAM that holds all three operands. 
    The spatial dataflow $s$ is fixed in accordance with the architecture.
    
    The total energy consumption of executing a layer is used as $V$.
    Experiments were run on a quad-core CPU @3.6GHz, and with $I=1000$, $\rho = 0.999$ and $T_0 = 0.05$.

    \vspace{-0.5em}
    \subsection{Experimental results}
    
    To assess the efficiency of the simulated annealing path of SALSA, we show the energy distribution of mappings using both SALSA and Timeloop (Fig.~\ref{fig:energydistrib}). Note that this energy distribution pattern is consistently found across layers of all studied DNNs. Compared to the random-pruned search of Timeloop, SALSA's simulated annealing energy distribution is centered on higher-quality states, providing better schedules in a shorter time.
    
    The stochastic nature of SALSA's simulated annealing motivates an exhaustive search on ResNet34 in order to study the capability of SALSA to consistently reach near-optimal schedules. We used LOMA to exhaustively find the best loop ordering for each unique layer of ResNet34, then we ran SALSA's simulated annealing engine 500 times per layer. We find that SALSA reaches the global optimum 99.9\% of the time. Even when SALSA does not find the global optimum, it still generates high-quality schedules, on average with $0.007$\% higher energy than the best mapping.
    
\begin{figure}[htbp]
    \vspace{-1em}
    \includegraphics[width=\columnwidth]{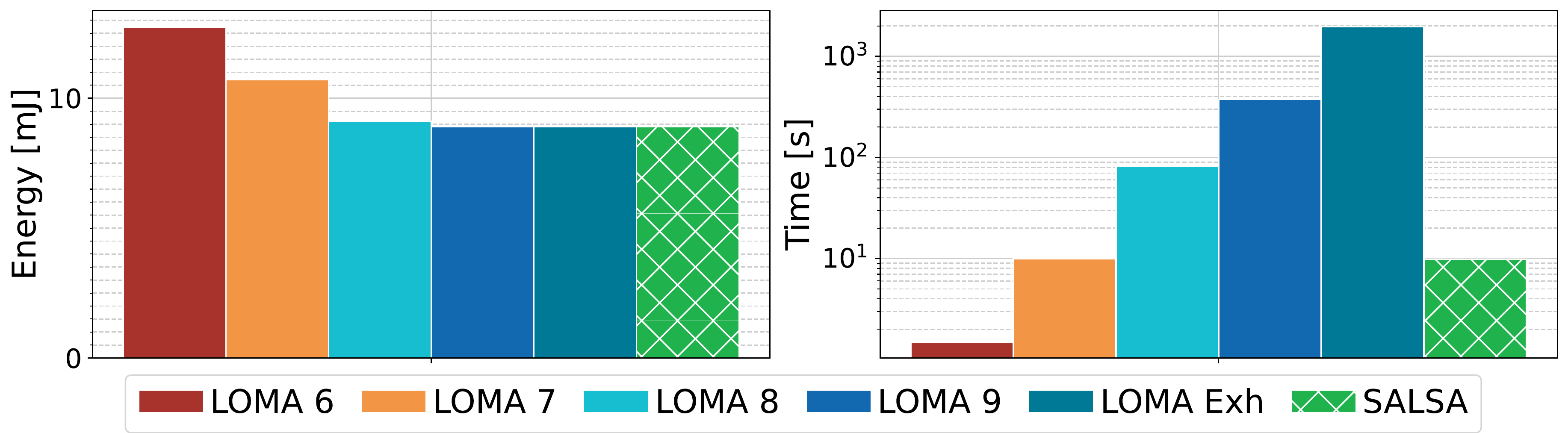}
    \vspace{-2em}
    \caption{Performances of SALSA against LOMA X on MobileNetV2 layer 3, X being the LPF limitation factor, shrinking the search space and trading mapping performances for search speed. The configuration of LOMA that does not constrain the search space is noted LOMA Exh (exhaustive).}
    \label{fig:salsa_loma}
    \vspace{-1em}
    \end{figure}

    We also compare SALSA against LOMA with various \textit{LPF Limits} (Fig.~\ref{fig:salsa_loma}). The \textit{LPF Limit} parameter indicates the maximum size of the orderings considered by LOMA, it limits the number of orderings to evaluate at the cost of the schedule's energy. We can clearly notice the trade-off between search time and energy between LOMA 6 and SALSA. Since the search for the optimal schedule is done offline, one would always favor lower energy rather than a reduction of a few seconds in the search time.
    
    Finally, we extensively benchmark SALSA against LOMA and Timeloop (Fig.~\ref{fig:unevenmerged}). We choose the LPF limitation factor of LOMA to get a similar search time to SALSA (see Fig.\ref{fig:salsa_loma}). In order to avoid a cost model bias, the schedule found by Timeloop's engine is evaluated using ZigZag's cost model. We notice that not all layers benefit from SALSA in the same way: all 3 search engines find similar energy schedules for simple layers (i.e., with fewer loops to permute). However, SALSA significantly outperforms LOMA and Timeloop for more complex layers with a bigger search space, leading to up to 50\% of energy reduction. Additionally, SALSA's search time is drastically lower than Timeloop's for every layer. Overall, SALSA improves the execution energy by 7.6$\%$, 11.9\%, and speed-up the search runtime by 1.7$\times$, 24$\times$, respectively.
    
\section{Conclusion}
\label{sec:conclusion}

    This paper presented SALSA: a dual-engine, rapid scheduler capable of finding optimal schedules of DNN layers onto an HW accelerator. The simulated annealing-based engine provides an efficient heuristic search guided by any desired performance metric and finds optimal mappings in a short and predictable time. 
    SALSA consistently finds better mappings than current SotA schedulers in a shorter time. It is deployed extensively on 5 DNNs: finding on average 7.6\% and 11.9\% better energy schedules while speeding up the search by a factor of 1.7$\times$ and 24$\times$ compared to LOMA and Timeloop, respectively.
    By significantly speeding up the process of extracting high-quality temporal mappings, SALSA paves the way for fast spatial unrolling and accelerator architecture search. SALSA is open-sourced and available at \cite{salsaopensource}.

\clearpage

\bibliographystyle{IEEEtran}
\bibliography{refs}

\end{document}